\documentclass[]{spie}
\usepackage{caption,subcaption}

\title{Laboratory characterization of a multi-photonic lantern optical waveguide using off-axis holography}

\author[a]{Aditya R. Sengupta}
\author[b]{Benjamin L. Gerard}
\author[b]{Dominic Sanchez}
\author[a]{Matthew DeMartino}
\author[a]{Rebecca Jensen-Clem}
\author[a]{Kevin Bundy}
\author[b]{Michael J. Messerly}
\author[b]{Paul Pax}
\author[a]{Daren Dillon}
\author[b]{Eric Strang}

\affil[a]{Department of Astronomy \& Astrophysics, University of California, Santa Cruz, CA 95064, USA}
\affil[b]{Lawrence Livermore National Laboratory, Livermore, CA 94550, USA}

\authorinfo{Further author information: (Send correspondence to A.S.)\\A.S.: E-mail: adityars@ucsc.edu}

\usepackage{preamble}

\begin{document} 
\maketitle

\begin{abstract}
    Photonic lanterns (PLs) are waveguides that convert multi-mode input light to single-mode outputs. Wavefront sensing (WFS) and spectroscopy using a PL have been demonstrated, but PL simulations and experiments show significant mismatches. For the WaveDriver project, a proposed Habitable Worlds Observatory pathfinder that uses a PL for WFS as well as for integral field spectroscopy, we manufactured an optical waveguide consisting of an array of seven 19-port PLs in one device. We present laboratory characterization of the individual PLs, consisting of measurements of the principal modes at each PL input using digital off-axis holography. We compare our mode measurements to simulations to assess the variation in the PL manufacturing process. We discuss expected WFS performance in the WaveDriver configuration.
\end{abstract}


\section{Introduction}

In order for the Habitable Worlds Observatory (HWO) to meet its picometer-level wavefront stability requirement, which is necessary in order to enable observations of Earth-like exoplanets at $10^{10}$ contrast, it requires active wavefront sensing (WFS) and control. WaveDriver is a mission concept in which a laser guide star spacecraft enables such active control \cite{GerardWaveDriver}. This needs to be paired with natural guide star (NGS) sensing for tip/tilt/focus control, and such sensing must also be done using a focal-plane wavefront sensor, i.e. a device that enables both wavefront sensing and science imaging, so that system performance is not dominated by non-common-path aberrations.

The photonic lantern (PL) is a waveguide that maps a multi-mode input to several single-mode outputs with high throughput \cite{Birks15}. PLs are wavefront sensors \cite{Norris20} whose outputs can also be used for multi-wavelength science imaging. The PL has been demonstrated on sky as a wavefront sensor \cite{Lin25,SenguptaOnSky} and for spectroscopy \cite{Vievard24}, making it a promising candidate for WaveDriver's NGS wavefront sensor. Simulations are necessary to demonstrate the PL's capabilities for WaveDriver. This can be done via methods like the beam propagation method \cite{lightbeam} and coupled-mode theory \cite{cbeam,LinCoupledModeTheory}. Previous simulation work has shown that the PL will enable a fainter limiting magnitude for HWO's NGS mode than the Zernike wavefront sensor \cite{GerardWaveDriver}.

However, due to inherent imprecision in the PL manufacturing process, the behavior of real PLs (i.e. the response at the single-mode end given a known input electric field at the multi-mode end) does not match their designs \cite{Rypalla24}. Experimental characterization of a manufactured lantern is necessary to bridge this gap. For WaveDriver concept development, an optical waveguide consisting of seven PLs in one device was manufactured at Lawrence Livermore National Laboratory, which was initially presented in Sengupta \textit{et al.} (2025)\cite{Sengupta25}. This device (henceforth `the Livermore lantern') is intended for use as a combined wavefront sensor and integral field spectrograph (IFS). Its function as an IFS is enabled by a unique fabrication approach that results in embedded geometric arrays of multiple lanterns, all in the same substrate. Instead of drawing each lantern one at a time, multiple preforms were drawn to an initial stage on LLNL's fiber tower. These were then assembled into a geometric array---separated by interstitial cladding rods---which constituted a second preform that was then drawn to the final desired dimensions. Fiber towers were previously used to make photonic lanterns \cite{Birks15,Leon-Saval05}, but were abandoned for this purpose in recent years. The Livermore lantern shows that fiber towers offer a path to mass-manufacture arrayed devices.

In this paper, we present laboratory characterization of this device using digital off-axis holography (OAH). This method has been used for PL characterization in other experiments: Xin \textit{et al.} (2024)\cite{XinJATIS} used OAH to characterize a 6-port mode-selective PL, and Taras \textit{et al.} (2026)\cite{Taras26} and Dobias \textit{et al.} (2026)\cite{Dobias26} used OAH to characterize 19-port standard PLs at multiple wavelengths. The characterization of the Livermore lantern presented in this paper consists of OAH measurements of the $19 \times 7 = 133$ single-mode ports. This work presents an opportunity to estimate the inherent variation in the PL manufacturing process, by characterizing multiple devices made to the same design specification under the same conditions.

The remainder of this paper is structured as follows. Section 2 presents the relevant theory of how photonic lanterns are mathematically modeled and how off-axis holography is relevant to constraining these models. Section 3 describes the design of the Livermore lantern and presents simulations of its mode structure and of OAH measurements. Section 4 describes the experimental setup. Section 5 presents the characterized model of the Livermore lantern and assesses its performance for wavefront sensing. Section 6 is a conclusion.

\section{Theory}

\subsection{Photonic lantern transfer matrices}

An $N$-port photonic lantern consists of a multi-mode end that supports $N$ modes (typically modeled as a step-index fiber), which adiabatically tapers to $N$ single-mode ports. At a single wavelength, PLs are fully characterized by a \textit{transfer matrix} (TM), which is an $N \times N$ complex-valued matrix where each element describes the coupling between one of the modes at the multi-mode end and one of the single-mode ports. Given a TM and a corresponding basis of modes, a lantern's response to an arbitrary electric field can be determined by decomposing the field into the basis of modes, multiplying the vector of basis coefficients by the TM, and taking a norm-squared to translate to the intensity at each single-mode port\cite{Lin22}. Although the TM of a real PL can be significantly different relative to its design, the TM is fixed once the PL is manufactured, and an identified TM can safely be used for performance analysis.

A PL can equivalently be characterized in terms of its \textit{principal modes}, which are the electric fields excited at the multi-mode end by injecting light into each of the single-mode ports (i.e. in the `backward' direction). The adiabaticity of the PL allows us to treat these as equivalent, i.e. exciting a principal mode at the multi-mode end would illuminate only one of the single-mode ports. The principal modes can be thought of as the basis in which the TM is the identity. Since illuminating each single-mode port and imaging the multi-mode side would yield only an intensity, rather than the full electric field necessary to identify the mode, we are required to additionally measure the phase component. The TM may alternatively be identified by illuminating the multi-mode end, varying the spatial structure of the input, and solving the resulting linear system or optimizing for the TM via gradient descent \cite{Kim24spectralcharacterization,Sengupta24,Romer2025}; however, this is strongly dependent on alignment quality and does not directly measure the relationship between the multi-mode end and each single-mode port in turn.

\subsection{Off-axis holography}

In the usual configuration, images only yield the intensity (i.e. the absolute value squared) of light, which is more fully described by a complex-valued electric field. Images cannot give us the phase component of the electric field, which is necessary in the context of PLs to measure the transfer matrix and make predictions about performance.

Digital OAH was first presented as a method of retrieving the full electric field of an unknown beam in by Cuche \textit{et al.} 2000 \cite{Cuche00}. OAH works by overlapping the unknown beam with a known reference beam to create interference; the resulting interferogram encodes the phase of the unknown beam. Appendix A provides a 1D derivation of the OAH concept showing that phase components are retrieved. Qualitatively, the reference beam being placed off-axis relative to the unknown beam, i.e. spatially separated in the pupil plane, causes interference in the focal plane. This interference can be backed out by taking a Fourier transform of the resulting image, which yields a convolution of the two pupil-plane beams (as well as an autoconvolution term that is masked out). Deconvolving the reference beam from this yields the full electric field of the unknown beam.

\begin{figure}[h!]
    \includegraphics[width=\textwidth]{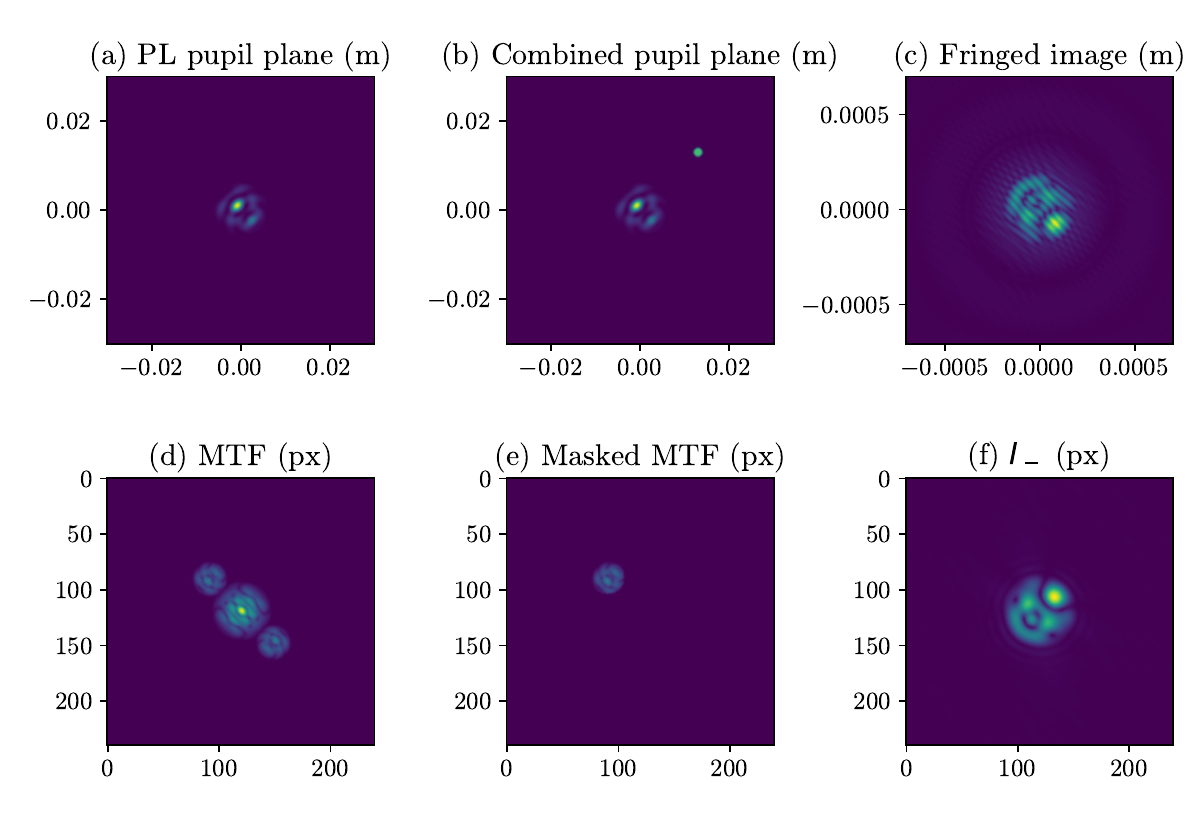}
    \caption{The intermediate steps in off-axis holography.}
    \label{fig:oah_steps}
\end{figure}

The relevant intermediate steps are as follows. These steps are the same as those used for the self-coherent camera as described in Gerard \textit{et al.} 2018\cite{Gerard18}. These steps are shown in Figure~\ref{fig:oah_steps}.

\begin{enumerate}[label=\alph*.]
    \item We obtain an on-axis image of the unknown beam, giving us its amplitude but not its phase.
    \item In the pupil plane, the unknown beam and reference beam are separated out. The required separation is at least 1.5 times the diameter of the unknown beam\cite{Galicher10}.
    \item In the focal plane, the two beams interfere to create a fringed image of the unknown beam.
    \item The Fourier transform of this fringed image is the optical transfer function (OTF). We plot its absolute value, referred to as the modulation transfer function (MTF).
    \item We mask out the side lobe to isolate only the component due to interference between the unknown beam and the reference beam.
    \item The inverse Fourier transform of the masked OTF is a complex-valued quantity we refer to as $I_-$. This is the product of the two focal-plane electric fields, which looks the same as (a) for a sufficiently small reference beam diameter.
\end{enumerate}

\section{Design and simulation of the Livermore lantern}

\subsection{Design}
\begin{figure}
    \begin{subfigure}{0.49\textwidth}
        \includegraphics[width=\linewidth]{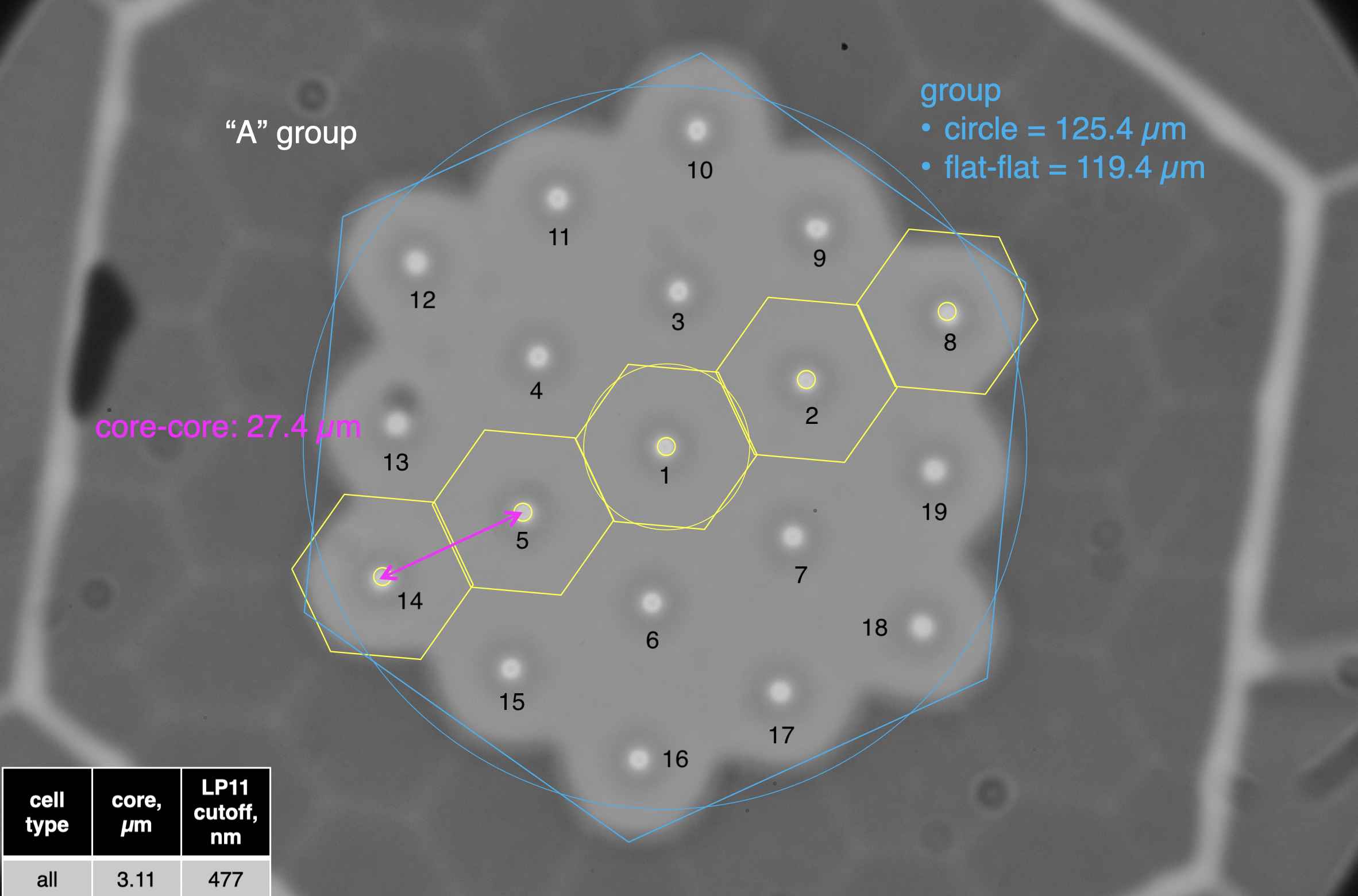}
        \caption{An `A' lantern.}
        \label{fig:microscope_a}
    \end{subfigure}\hspace*{\fill}
    \begin{subfigure}{0.49\textwidth}
        \includegraphics[width=\linewidth]{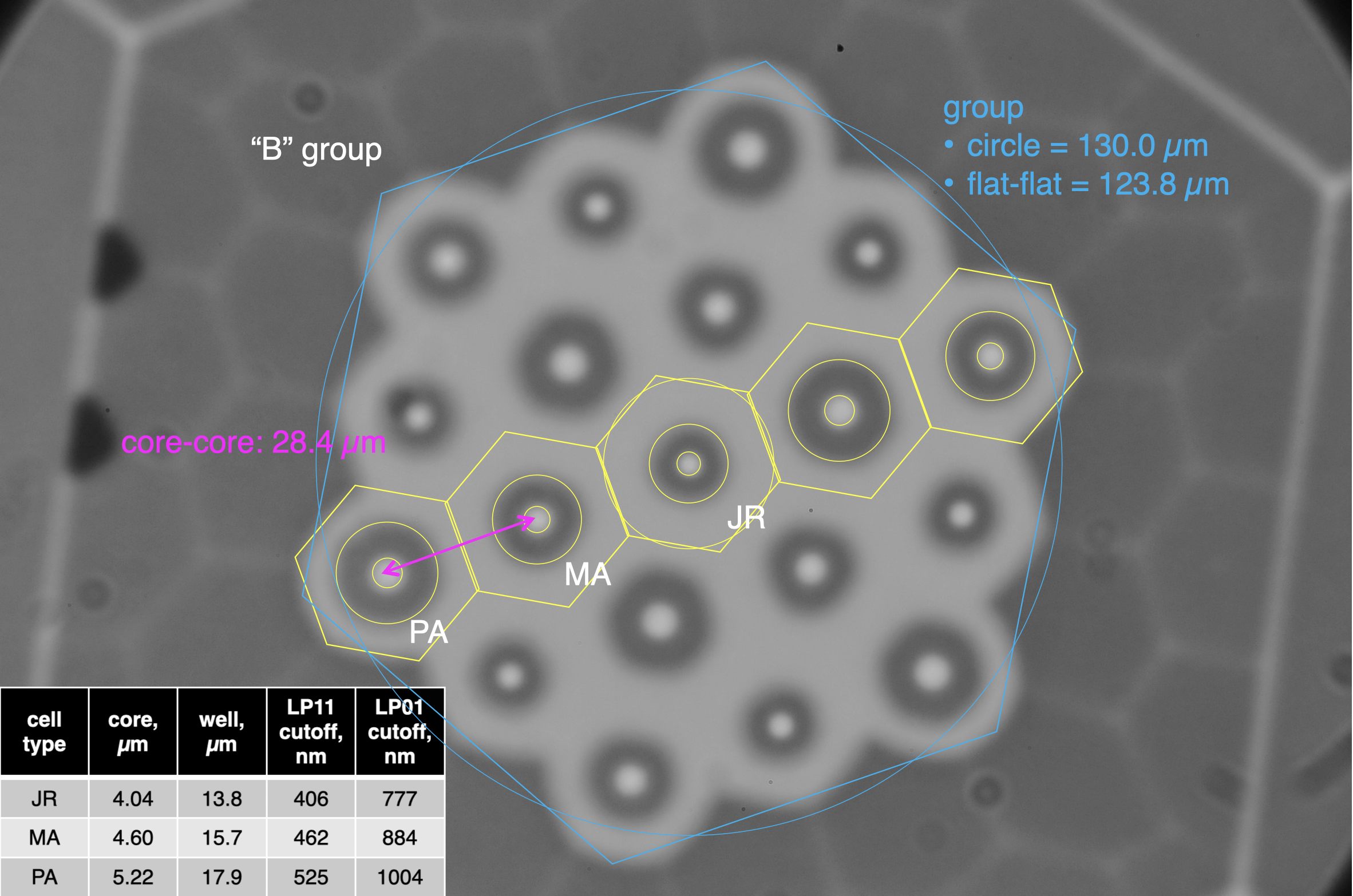}
        \caption{A `B' lantern.}
        \label{fig:microscope_b}
    \end{subfigure}
    \caption{Microscope images of the two lantern designs.}
    \label{fig:microscope_lanterns}
\end{figure}

The device being characterized in this work consists of seven PLs within one monolithic piece of glass, each of which has 19 ports. Four of the lanterns have standard designs, i.e. all ports have the same size and are positioned in a uniform hexagonal pattern; these are referred to as the `A' design. The remaining three have slight variations in their port sizes, alternating between smaller and larger ports within each ring, and feature a depressed-well structure; these are referred to as the `B' design. The PL was fabricated in a multi-step process using two different single-mode pre-forms for the two designs. The A design is from OFS and the B design is from Weatherford (Preform 190916-1). The B design preform uses two different fiber core materials and with three different core sizes throughout the 19 cores (6 large, 6 medium, 7 small). Both bundles are designed for a 633nm central wavelength, but the B design provides a potentially larger spectral range over which single mode behavior exists but at the expense of decreased throughput. The design parameters are listed in Table~\ref{tab:design_parameters}, and labeled microscope images of the two designs are shown in Figure~\ref{fig:microscope_lanterns}. The microscope image of the A design also shows the port index convention used throughout this work. We later label the individual lanterns based on their relative position in the overall lattice, with the labels `central', `lefta', `righta', and `bottoma' for the A lanterns and `bottomb', `leftb' and `topb' for the B lanterns.

\begin{table}
    \centering
    \begin{tabular}{|c|c|}
        \hline
        \textbf{Common parameters} & \\
        Taper factor & 7.14 \\
        Taper length & 1cm\\
        Numerical aperture & 0.117\\
        Core refractive index & 1.4631 \\
        Cladding refractive index & 1.4584 \\
        Jacket refractive index & 1.4530 \\
        Cladding diameter & 125.4 $\mu$m \\
        \textbf{`A' design parameters} & \\
        Core diameter & 3.11 $\mu$m \\
        Core-to-core spacing & 27.4 $\mu$m \\
        \textbf{`B' design parameters} & \\
        Core diameter, JR & 4.04 $\mu$m \\
        Core diameter, MA & 4.60 $\mu$m \\
        Core diameter, PA & 5.22 $\mu$m \\
        Core-to-core spacing & 28.4 $\mu$m\\\hline
    \end{tabular}
    \caption{Design parameters for the Livermore lantern.}
    \label{tab:design_parameters}
\end{table}

\subsection{Simulation}
\begin{figure}[h!]
    \begin{subfigure}{0.49\textwidth}
        \includegraphics[width=\linewidth]{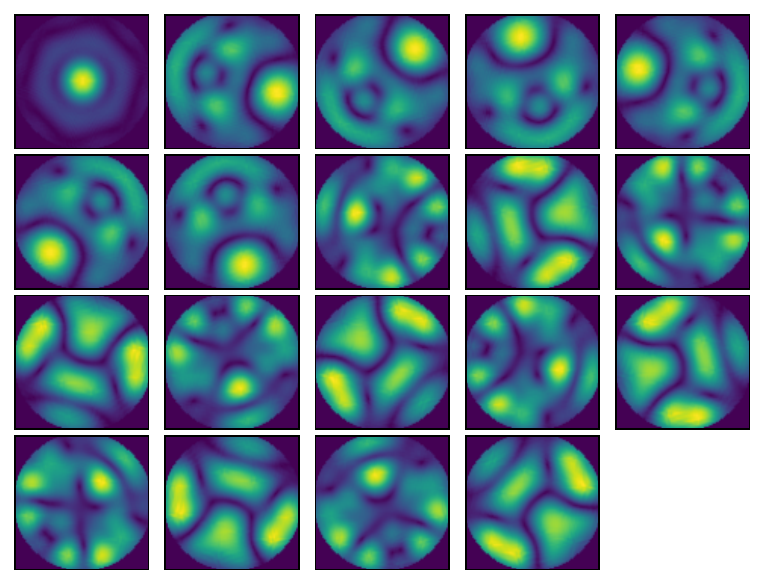}
        \caption{Livermore-A principal modes, amplitude}
        \label{fig:pl_a_abs}
    \end{subfigure}\hspace*{\fill}
    \begin{subfigure}{0.49\textwidth}
        \includegraphics[width=\linewidth]{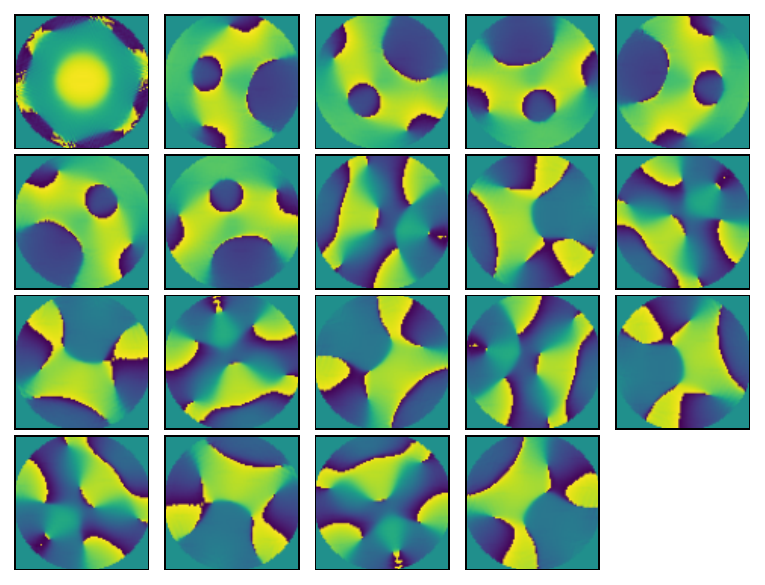}
        \caption{Livermore-A principal modes, phase}
        \label{fig:pl_a_angle}
    \end{subfigure}
    \caption{Livermore-A simulated principal modes.}
    \label{fig:pl_a}
\end{figure}

\begin{figure}[h!]
    \begin{subfigure}{0.49\textwidth}
        \includegraphics[width=\linewidth]{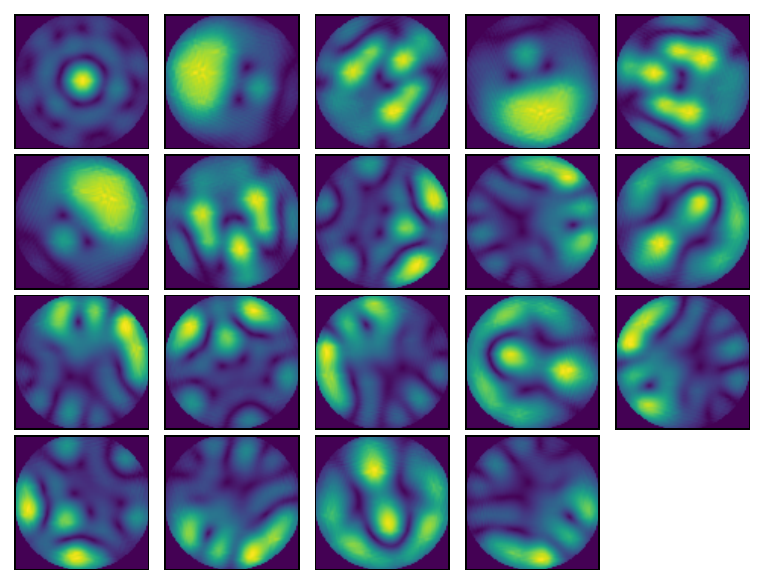}
        \caption{Livermore-B principal modes, amplitude}
        \label{fig:pl_b_abs}
    \end{subfigure}\hspace*{\fill}
    \begin{subfigure}{0.49\textwidth}
        \includegraphics[width=\linewidth]{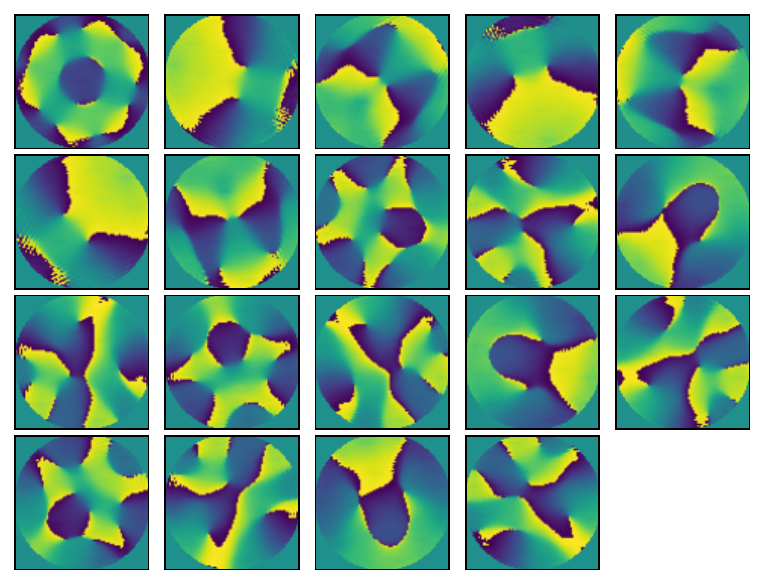}
        \caption{Livermore-B principal modes, phase}
        \label{fig:pl_b_angle}
    \end{subfigure}
    \caption{Livermore-B simulated principal modes.}
    \label{fig:pl_b}
\end{figure}

We simulated the two lantern designs using the \textit{lightbeam} Python package\cite{lightbeam}. In these simulations, we illuminate each single-mode port in turn and compute the corresponding multi-mode patterns, meaning a direct outcome of the simulation is the full electric field of each mode, i.e. the same quantity we are trying to measure. Note that the B design simulation does not include the depressed well. We use a grid resolution of 0.2 $\mu$m in the $x$ and $y$ directions, 25 $\mu$m in the z direction, an extent in $x$ and
$y$ of 150$\mu$m, and 8 grid cells on each boundary in $x$ and $y$ used as perfectly matched layers. Figures~\ref{fig:pl_a} and~\ref{fig:pl_b} show the amplitude and phase of each mode for the A and B designs respectively.

\begin{figure}[h!]
    \begin{subfigure}{0.49\textwidth}
        \includegraphics[width=\linewidth]{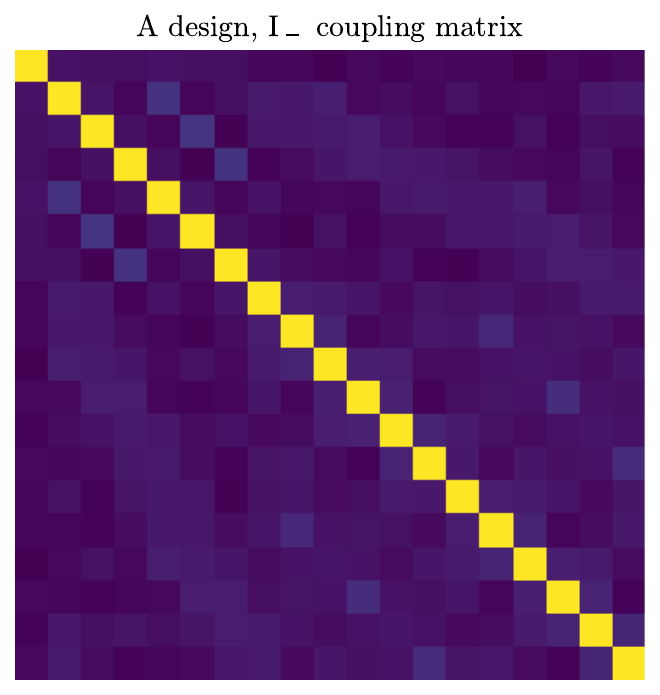}
        \caption{`A' design.}
        \label{fig:iminus_coupling_a}
    \end{subfigure}\hspace*{\fill}
    \begin{subfigure}{0.49\textwidth}
        \includegraphics[width=\linewidth]{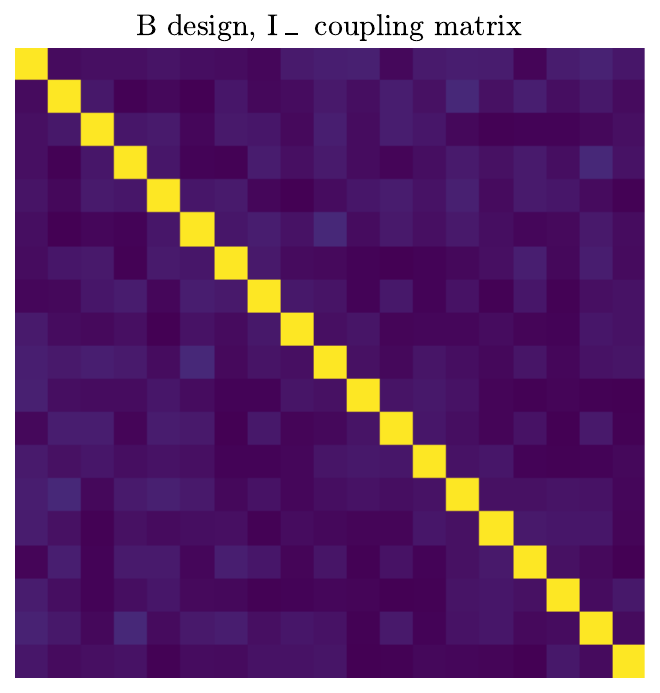}
        \caption{`B' design.}
        \label{fig:iminus_coupling_b}
    \end{subfigure}
    \caption{$I_-$ coupling matrices.}
    \label{fig:iminus_coupling}
\end{figure}

We use the $I_-$ coupling matrix, consisting of the overlap integrals of each pair of $I_-$s within a lantern, as an observable to directly compare the manufactured lanterns to their design. Figures~\ref{fig:iminus_coupling_a} and~\ref{fig:iminus_coupling_b} show these coupling matrices for the A and B designs respectively. Note that these matrices are not the same as the transfer matrices. The coupling matrices show correlations between ports, i.e. they show which ports tend to respond together. This is a relevant feature of standard or near-standard lanterns for wavefront sensing, as multiple ports responding to an aberration creates a clearer signal that is more robust to noise. Although most correlations are relatively low, we note a cluster in the upper left of the A coupling matrix, corresponding to the central port and inner ring, and banded structures throughout corresponding to port adjacencies.  

\begin{figure}[hbtp!]
    \begin{subfigure}{0.49\textwidth}
        \includegraphics[width=\linewidth]{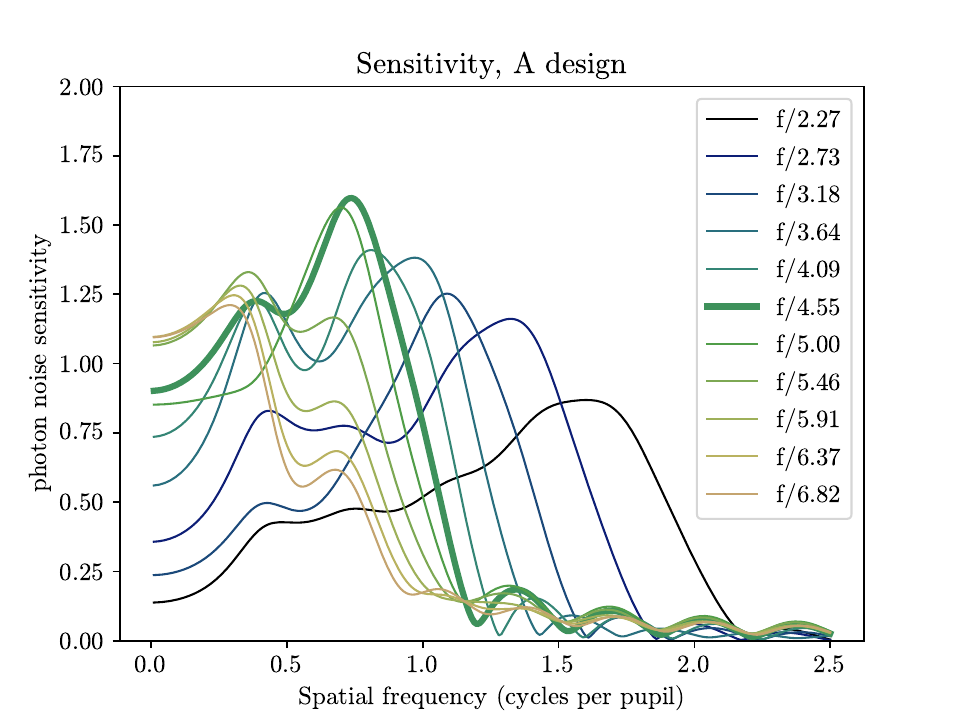}
        \caption{`A' design.}
        \label{fig:livermore_a_photon_sensitivity}
    \end{subfigure}\hspace*{\fill}
    \begin{subfigure}{0.49\textwidth}
        \includegraphics[width=\linewidth]{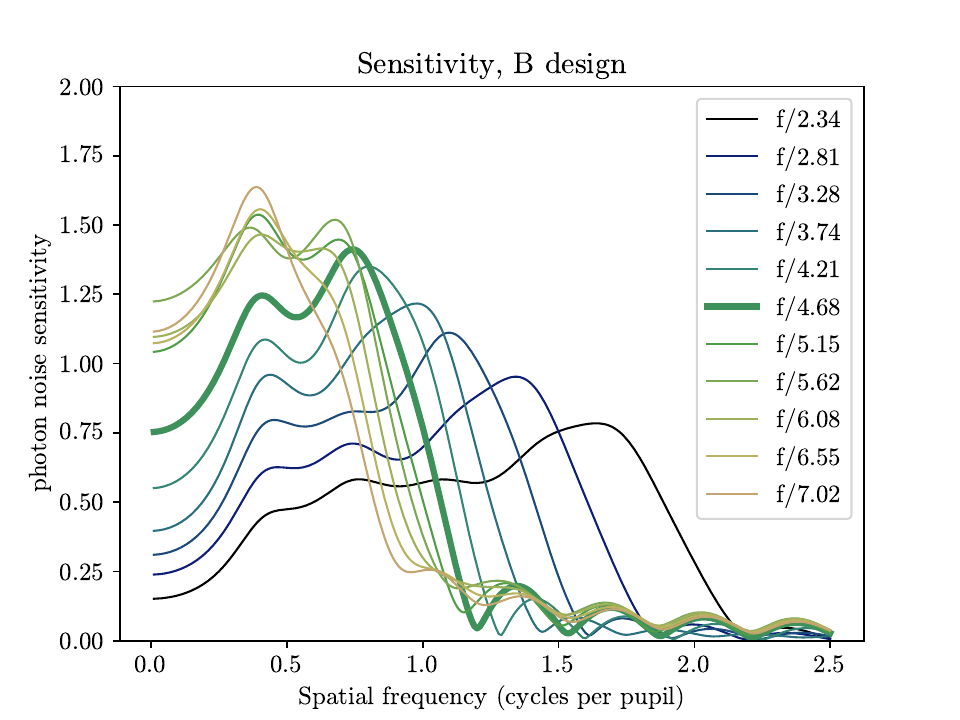}
        \caption{`B' design.}
        \label{fig:livermore_b_photon_sensitivity}
    \end{subfigure}
    \caption{Photon noise sensitivity curves. The f-numbers are linearly spaced around the one that provides the highest throughput (which is determined in simulation and is at $1 / (2 \times \text{NA})$), which is shown with a thicker line.}
    \label{fig:photon_sensitivity_sim}
\end{figure}

Manufactured lanterns being significantly different than their designs does not necessarily imply their WFS performance is significantly different. Therefore, we also compare WFS sensitivity with respect to photon noise, for sine-wave aberrations as a function of spatial frequency in cycles per pupil. The sensitivity metric is described in Chambouleyron \textit{et al.} 2023\cite{Chambouleyron23}, and the implementation for photonic lanterns is described in more detail in Sengupta \textit{et al.} in these proceedings\cite{Sengupta26}. We plot sensitivity at a range of f-numbers to show the trade space of sensitivities vs. upper limits on spatial frequency that is achievable. A more complete characterization of the role of lantern sensitivity in AO error budgets is out of scope for this work; here, several curves are shown in order to demonstrate the range of capabilities that the Livermore lantern could provide.

\section{Experimental setup}

\begin{figure}[h!]
    \includegraphics[width=\textwidth]{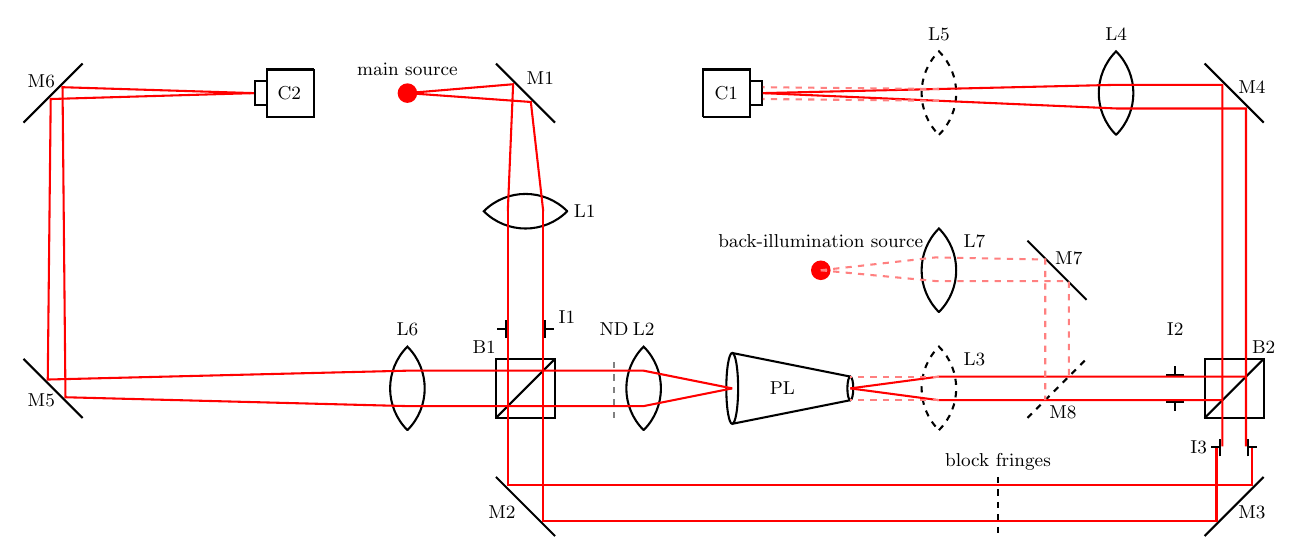}
    \caption{The optical layout for off-axis holography of the Livermore lantern. The primary optical path is in solid red, and auxiliary paths for back-reflection imaging and pupil-plane imaging that are enabled by flipping optics in and out are in dashed red. The optics that can be flipped in and out to switch between these modes are in dashed black.}
    \label{fig:oah_diagram}
\end{figure}

\begin{figure}[h!]
    \includegraphics[width=\textwidth]{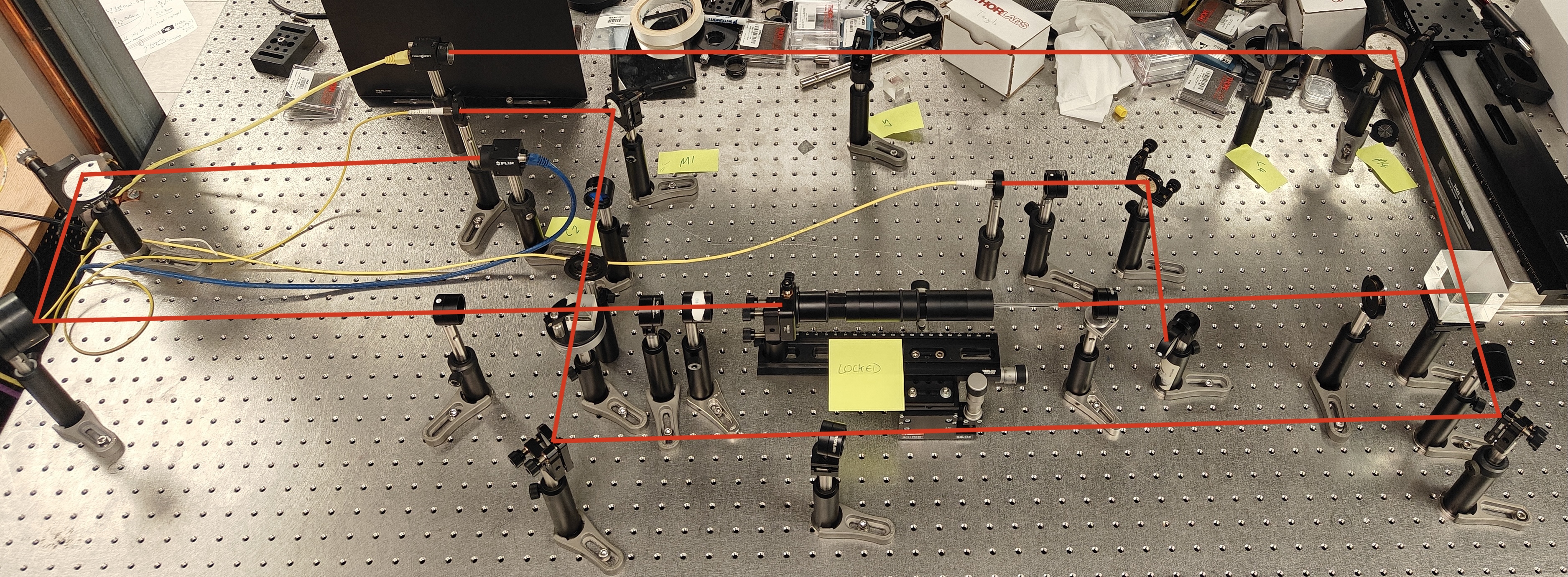}
    \caption{The laboratory setup for off-axis holography of the Livermore lantern. The optical path is indicated in red.}
    \label{fig:oah_bench}
\end{figure}

We built an optical setup in the UCSC Laboratory for Adaptive Optics to conduct OAH measurements. The layout is shown in Figures~\ref{fig:oah_diagram} and~\ref{fig:oah_bench}, and the full set of components is described in Table~\ref{tab:components}. Along the primary optical path, the beam is collimated and split: one component is brought to a focus on the lantern's single-mode end, and the output from the multi-mode end is recollimated and cropped to isolate only the multi-mode core and exclude the wider cladding. Note that this is the reverse orientation from the multi-mode to single-mode order in which lanterns are normally set up. The other component is reflected, cropped, and shifted off-axis in order to create interference. The two beams are then recombined and reimaged. To accurately place the reference beam off-axis, L5 is positioned in a telecentric imaging configuration such that it images I2 and I3 onto C1. L5 is placed on a flip mount to allow switching between pupil-plane and focal-plane imaging mode. 

\begin{table}[h]
    \begin{tabular}{|p{0.22\textwidth}|p{0.12\textwidth}|p{0.58\textwidth}|}
        \hline
        \textbf{Label} & \textbf{Type} & \textbf{Description} \\\hline
        main source & Laser & HeNe laser, primary source for OAH measurements \\
        M1 & Mirror & Mirror with tip/tilt knobs, sets constant beam height \\
        L1 & Lens & $f = 300$mm lens, collimates the main source \\
        B1 & Beam cube & Splits into lantern path and reference path \\
        I1 & Iris & Sets the PSF size on the lantern entrance \\
        ND & Filter & Neutral-density filter on a flip mount, to optimize fringes \\
        L2 & Lens & $f = 60$mm lens, focuses beam at the lantern's single-mode end; on flip mount \\
        PL & Lantern & The Livermore lantern \\
        L3 & Lens & $f = 50$mm lens to collimate the post-lantern beam \\ 
        I2 & Iris & Crops the lantern beam to isolate the multimode core \\ 
        M2 & Mirror & Redirects the reference beam \\
        block fringes & Mask & Opaque mask on a flip mount to obtain unfringed images \\
        M3 & Mirror & Redirects the reference beam \\
        I3 & Pinhole & $400\mu$m pinhole, crops the reference beam and is moved off-axis to meet the interference condition \\
        B2 & Beam cube & Recombines the two beams; 2'' to fit both beams off-axis \\
        M4 & Mirror & Redirects the combined beam \\
        L4 & Lens & $f = 1000$mm lens for final imaging \\
        L5 & Lens & $f = 400$mm lens for telecentric pupil-plane imaging \\
        C1 & Camera & Final imaging camera, Blackfly U3-23S6M-C, 1 px = 5.86 $\mu$m \\ 
        L6 & Lens & $f = 1000$mm lens for back-illuminated imaging \\
        M5 & Mirror & Redirects the back-illuminated beam \\
        M6 & Mirror & Redirects the back-illuminated beam \\
        C2 & Camera & Back-illumination camera, Blackfly U3-23S6M-C, 1 px = 5.86 $\mu$m \\
        back-illumination source & Laser & HeNe laser to illuminate the multi-mode end of the lantern \\
        L7 & Lens & $f = 60$mm lens, collimates the back-illumination source \\
        M7 & Mirror & Redirects the back-illumination beam \\
        M8 & Mirror & Redirects the back-illumination beam; on a flip mount\\
        C2 & Camera & Back-illumination camera \\\hline
    \end{tabular}
    \caption{The optical components used in this experimental setup.}
    \label{tab:components}
\end{table}

In order to position the PSF at the lantern's single-mode end on a particular core, the PL is mounted on a five-axis stage, and B1 has knobs for tip/tilt control. Targeting particular cores in a known order additionally requires imaging of the single-mode end with the PSF overlaid. Therefore, an additional source is injected into the multi-mode end, so that the single-mode end is uniformly illuminated and the cores can be identified. The beam from this source is directed into the PL's multi-mode end using a flip mirror (M8), which is flipped out of the beam when collecting OAH images. L3 is also on a flip mount, so that it can be flipped in for the main imaging mode and flipped out to allow for a flat field for the back-illuminated imaging mode. The resulting image from illuminating the single-mode end in this way is captured at C2. Additionally, the PSF formed by L2 causes a back-reflection that is also captured at C2, allowing us to move the PSF and lantern relative to one another to target individual cores. In order to clearly see the structure of the single-mode end without also saturating the PSF, a more powerful laser was required for the back-illumination path than the main path. After initial alignment, the tip/tilt of the PL was not moved in order to keep the post-lantern beam in approximately the same position. However, the multi-mode end of the lantern did move slightly with x-y translation of the stage, and this movement was magnified by a factor of 20 (which is the ratio of the focal lengths of L4 and L3) at C1. We compensate for this by adjusting the tip/tilt of B2 slightly per measurement as necessary, and by creating a large interference envelope using a small diameter for I3 (400 $\mu$m).

\begin{figure}[t!]
    \includegraphics[width=\textwidth]{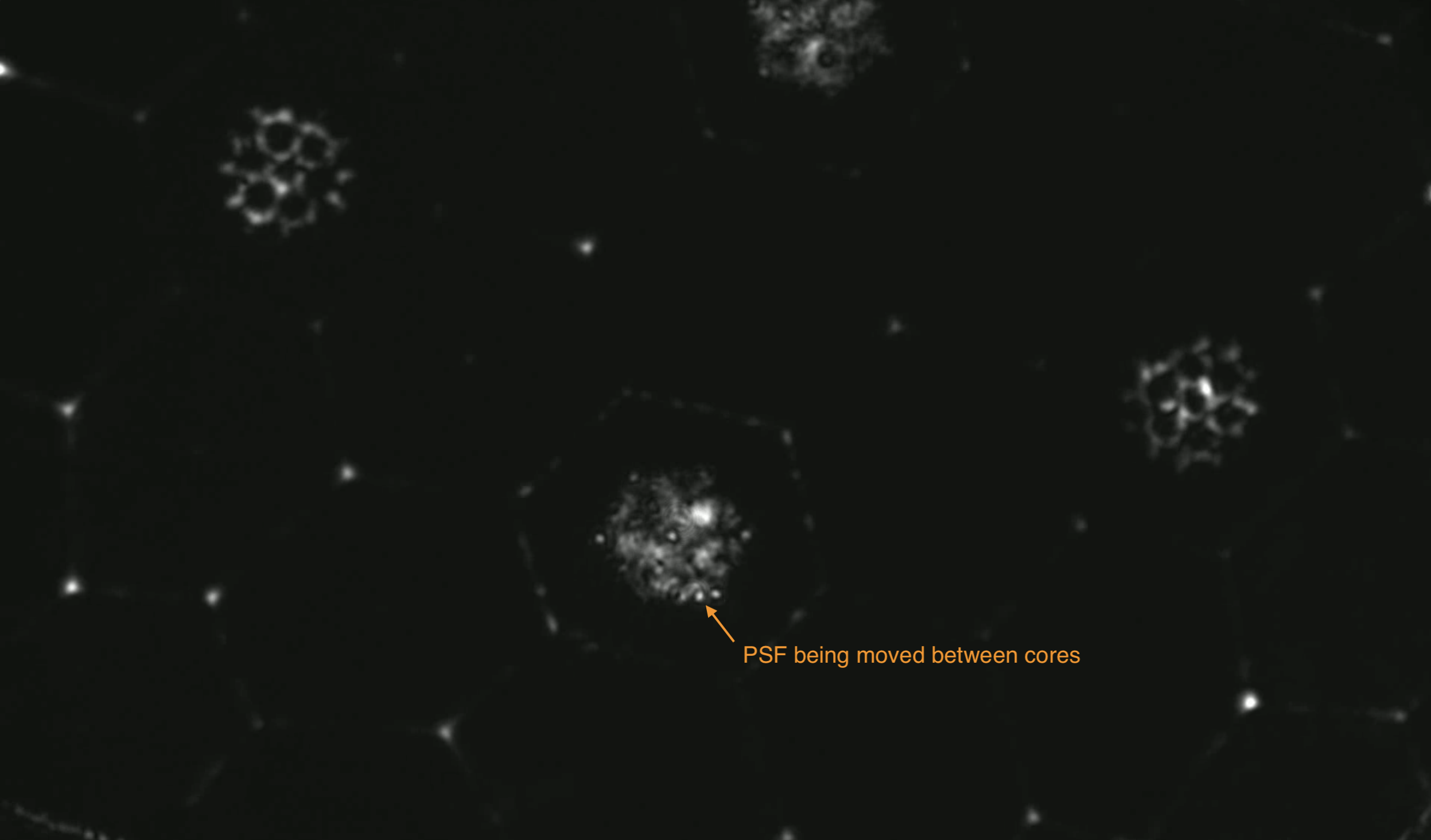}
    \caption{A representative frame from C2, showing several lanterns and the PSF being moved between the cores.}
    \label{fig:backillumination}
\end{figure}

Figure~\ref{fig:backillumination} shows a representative frame from C2. The acquisition process is as follows. By moving B1 and the PL mount, we move the PSF onto a particular core as seen at C2. We note that the cores for the B design cannot be clearly resolved at C2, despite being at similar radii and core-to-core separations as the A designs. However, the cores can still be targeted by approximately positioning the PSF on C2 and then moving slightly until the brightness on C1 is locally maximized. Once this is done, we adjust B2 as needed to move the lantern beam within the interference envelope while meeting the separation criterion. If fringes are not clearly visible on the image, the exposure time and the neutral-density filter in the lantern path are adjusted to obtain an image with suitable contrast between the paths. We measure a fringed image and plot its MTF to ensure the side lobes are well separated. We further take an unfringed image and a dark image.

We dark-subtract the fringed images and take cutouts to correct for position shifts on the detector. Cutouts were initially centered around the brightest pixel in the image and were refined by hand to include the full mode images. We note that position offsets in this hand-refinement process are a potential source of error.

\begin{figure}[h!]
    \includegraphics[width=\textwidth]{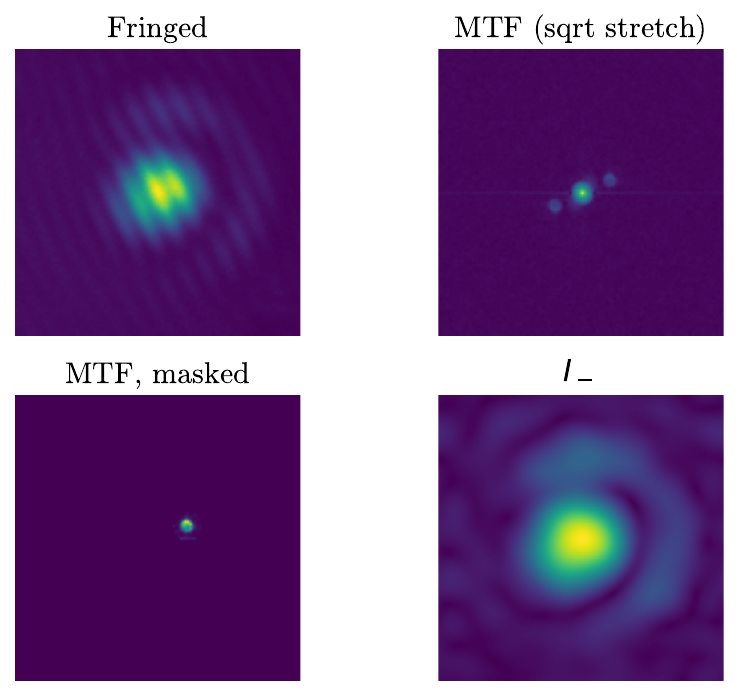}
    \caption{The image reduction steps for one port.}
    \label{fig:example_reduction}
\end{figure}

Figure~\ref{fig:example_reduction} shows the intermediate reduction steps (c-f from Figure~\ref{fig:oah_steps}) for port 1 of the central lantern. We take as our identified mode the inverse Fourier transform of the masked OTF; this is actually the product of the mode and the PSF from the reference beam, but since the diameter of the reference beam is relatively small (400 $\mu$m) and the focal length at C1 is large (1m), the resulting PSF core has a size of $f\lambda/D = (1\text{m}) \times (0.633\mu\text{m}) / (400 \mu\text{m}) = 1.5825 \text{mm}$, or approximately 270 px. This is significantly larger than the mode shape cutouts, which were 120 px across, so the variation across the mode shape due to the reference beam PSF is negligible other than a factor of overall intensity.

\section{Results and assessment of WFS performance}

\begin{figure}[h!]
    \includegraphics[width=\textwidth]{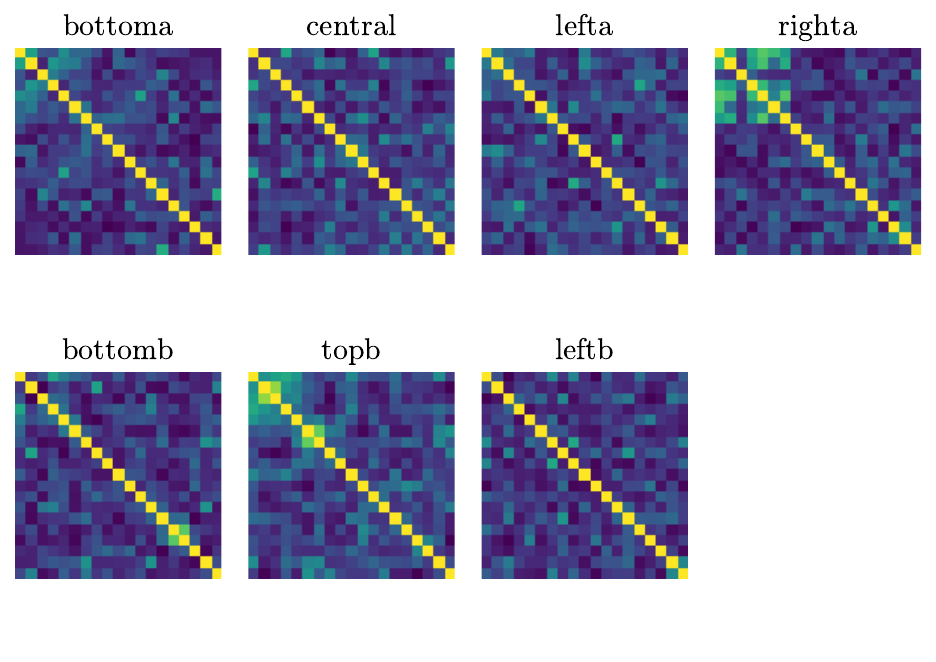}
    \caption{The $I_-$ coupling matrices for all seven lanterns.}
    \label{fig:experimental_coupling_matrices}
\end{figure}

Figure~\ref{fig:experimental_coupling_matrices} shows the $I_-$ coupling matrices for each lantern. We note in both that while some lanterns have stronger upper-left correlations, as seen in simulations, others do not. Further, no banded diagonal structures are clearly apparent, although there is significant short-range structure, where three or four adjacent ports tend to respond together (e.g. in the `topb' design.)

We compute WFS sensitivity to photon noise for the identified lanterns using the retrieved modes instead of the simulated ones. Since all of the fringed images were taken at different exposure times and ND filter settings to ensure fringe visibility, we do not normalize modes based on experimental data; instead, we normalize all modes within a lantern to the overall peak throughput in simulations (0.946 for A, 0.905 for B.)

\begin{figure}[h!]
    \begin{subfigure}{0.49\textwidth}
        \includegraphics[width=\linewidth]{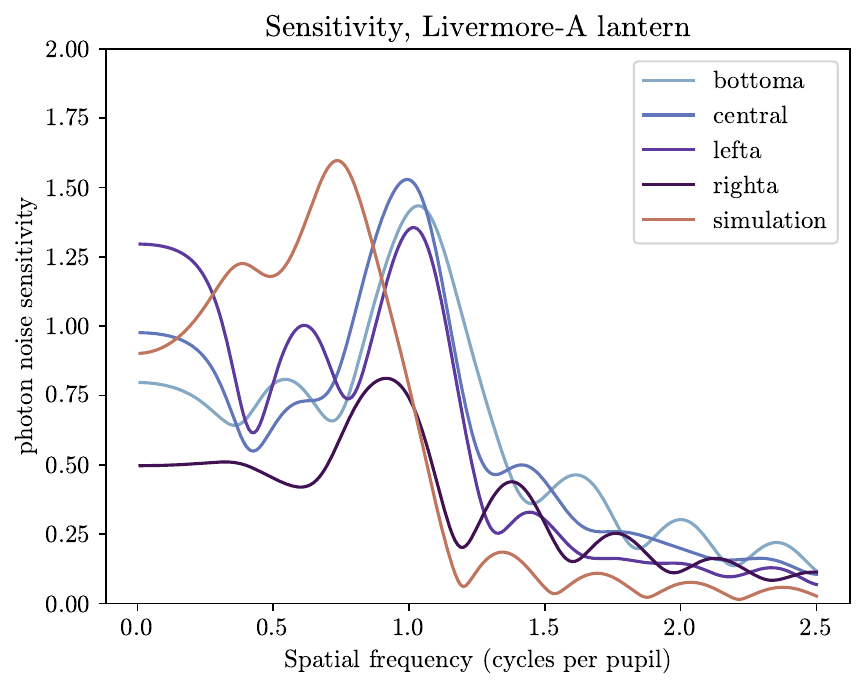}
        \caption{`A' design.}
        \label{fig:livermore_a_photon_sensitivity_experimental}
    \end{subfigure}\hspace*{\fill}
    \begin{subfigure}{0.49\textwidth}
        \includegraphics[width=\linewidth]{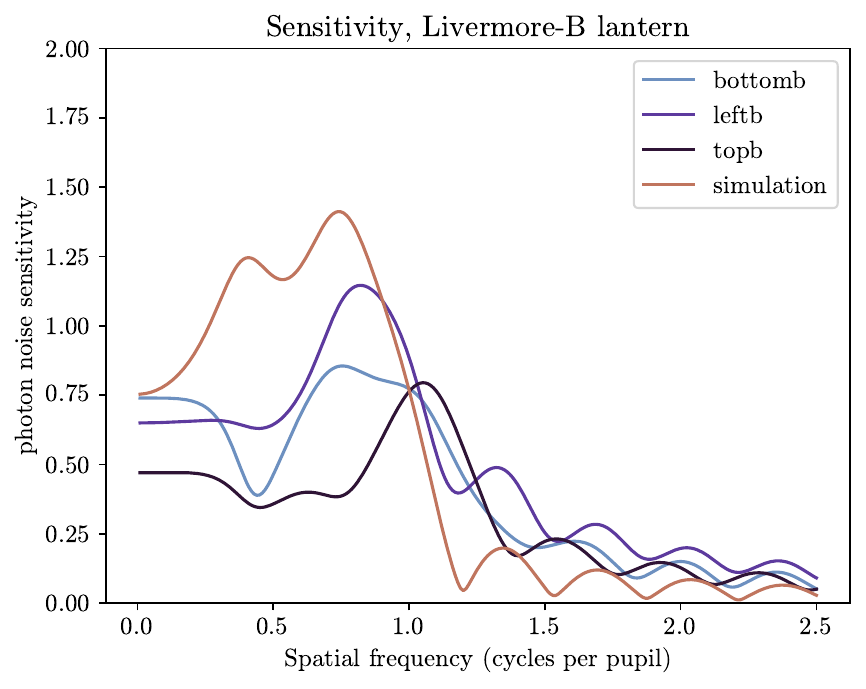}
        \caption{`B' design.}
        \label{fig:livermore_b_photon_sensitivity_experimental}
    \end{subfigure}
    \caption{The photon noise sensitivity of the characterized lanterns relative to the simulations from Figure~\ref{fig:photon_sensitivity_sim}. All simulations are run at the f-number that showed the peak throughput for that design in simulation (f/4.55 for A, f/4.68 for B.)}
    \label{fig:photon_sensitivity_exp}
\end{figure}

Figure~\ref{fig:photon_sensitivity_exp} shows the resulting sensitivity curves. All characterized lanterns show peaks and troughs at roughly the same spatial frequencies, with the exception of the `leftb' lantern relative to the other two B designs. This suggests the peaks are a feature of shared geometry, e.g. having the same port-to-port spacing or the same multimode input size. In both the A and B lanterns, the main peak is located slightly rightward (i.e. at a slightly higher spatial frequency) relative to the simulated peak. This may imply the true f-number at which the peak throughput is achieved is slightly smaller for the lanterns as manufactured relative to the simulations, as the smaller f-numbers show peaks at higher spatial frequencies due to more Airy rings coupling into the same space for a smaller PSF core.

There is noticeable variation in the sensitivity values; these differences may be due to factors that are distinct from lantern to lantern, such as the internal structure of the taper. Three of the A lanterns show very consistent sensitivity curves relative to one another and, other than the peak offset, relative to the simulation. The exception is the `righta' lantern, which is also the only A lantern to have shown a strong upper-left cluster in its $I_-$ coupling matrix, although a causative relationship between these two results is unclear, since high correlations in the inner ring should benefit sensitivity. We note that a similar cluster is visible in the `topb' lantern's coupling matrix, which also shows relatively low photon noise sensitivity. The B lanterns more consistently underperform the simulation; a possible cause of this is not simulating the depressed well, which may have led to an overestimate of sensitivity as the throughput reduction may have been underestimated. We note that the characterized A lanterns are mostly more sensitive than the B lanterns, as expected from the initial design intent.

\section{Conclusion}

We have characterized a novel type of optical waveguide consisting of seven photonic lanterns made on one device via digital off-axis holography. We identified the principal modes of each lantern and used these modes to characterize lantern-to-lantern differences and the suitability of each lantern for wavefront sensing.

This work presents one of the first tests of the consistency of photonic lantern manufacturing: all seven lanterns were made within the same overall device, with the same materials and under the same laboratory conditions. We find that variation in the manufacturing conditions did not in this case qualitatively change the lantern's behavior, e.g. prevent light from being transmitted through a port entirely or make the lantern blind to a particular aberration mode. There is significant variation in the delivered lanterns that can be observed via the $I_-$ coupling matrices, and in some cases this affects WFS performance as seen in estimates of photon noise sensitivity. Nevertheless, we find that each lantern shows some WFS sensitivity to all the modes considered, i.e. all the modes we would expect the lantern to be sensitive to, based on simulations and on lantern geometry (port count, port layout, number of Airy rings being coupled in at the peak throughput.)

The intended use for this device is as a joint wavefront sensor and integral field spectrograph. Single-wavelength OAH measurements are sufficient to constrain this lantern's WFS capabilities, as monochromatic WFS performance of a lantern in the WaveDriver configuration was previously found to closely match the dispersed multi-wavelength case in terms of photon noise sensitivity\cite{GerardWaveDriver}. We therefore find that this device is suitable as a WFS for this dual-purpose configuration. Future work will demonstrate the combined WFS and spectroscopic capabilities of this lantern in more detail. 


\acknowledgments     
A.S. thanks Philip Hinz, Jordan Diaz, and Parth Nobel for valuable discussions, and Deno Stelter and Isabel Kain for help with the experimental setup. This work was performed under the auspices of the U.S. Department of Energy by Lawrence Livermore National Laboratory under Contract DE-AC52-07NA27344. This document was prepared as an account of work sponsored by an agency of the United States government. Neither the United States government nor Lawrence Livermore National Security, LLC, nor any of their employees makes any warranty, expressed or implied, or assumes any legal liability or responsibility for the accuracy, completeness, or usefulness of any information, apparatus, product, or process disclosed, or represents that its use would not infringe privately owned rights. Reference herein to any specific commercial product, process, or service by trade name, trademark, manufacturer,or otherwise does not necessarily constitute or imply its endorsement recommendation, or favoring by the United States government or Lawrence Livermore National Security, LLC. The views and opinions of authors expressed herein do not necessarily state or reflect those of the United States government or Lawrence Livermore National Security, LLC and shall not be used for advertising or product endorsement purposes. The document number is LLNL-PROC-2020380.

\appendix
\section{1D OAH derivation}

We present a 1D derivation demonstrating that OAH finds the full electric field of the unknown beam. Over a 1D domain parameterized by $x$, suppose we have a test function $E(x)$ whose support is $\left[-\frac{A}{2}, \frac{A}{2}\right]$ but that is otherwise unknown, and a reference function $R(x) = \text{rect}\left(\frac{x - C}{B}\right)$, where 

\begin{align}
    \text{rect}(x) = \begin{cases} 1 & |x| \leq 1/2 \\ 0 & |x| > 1 / 2 \end{cases}
\end{align}

and where we choose support $\left[\frac{C - B}{2}, \frac{C + B}{2}\right]$ for our reference function. We choose $C > \frac{A + B}{2}$ to ensure the functions do not overlap.

Using the Fourier transform identities

\begin{align}
    g(x) \ast h(x) &\leftrightarrow G(f) H(f)\\
    g^*(-x) &\leftrightarrow G^*(x)
\end{align}

where $\ast$ is the convolution operator, we can say

\begin{align}
    \mathcal{F}^{-1} \left(\abs{\mathcal{F}(g(x))}^2\right)= g^*(-x) \ast g(x).
\end{align}

If we let $g(x) = E(x) + R(x)$, this gives us

\begin{align}
    \mathcal{F}^{-1} \left(\abs{\mathcal{F}(E(x) + R(x))}^2\right) = E^*(-x) \ast E(x) + R(-x) \ast E(x) + E^*(-x) \ast R(x) + R(-x) \ast R(x).
\end{align}

Note that we have $R^*(x) = R(x)$ as it is a real function. The left-hand side of this operation is analogous to adding the unknown and reference beams together in the pupil plane, propagating to the focal plane via a Fourier transform, measuring the intensity by taking a norm-squared, and taking an inverse Fourier transform to get the optical transfer function (OTF). This has four terms, but only three visible components, because $E^*(-x) \ast E(x)$ and $R(-x) \ast R(x)$ are both centered on $x = 0$ and do not contain the relevant phase information. The remaining two visible components (the `side lobes') have support from $\left[-C - \frac{A + B}{2}, -C + \frac{A + B}{2}\right]$ on the negative end and $\left[C - \frac{A + B}{2}, C + \frac{A + B}{2}\right]$ on the positive end. The side lobes do not overlap with the central component due to our choice of $C$ sufficiently large. Both components carry the same information, so without loss of generality we mask out everything but a side lobe and we take 

\begin{align}
    OTF(x) = R(-x) \ast E(x).
\end{align}

To generate $I_-$, we take a further Fourier transform of the OTF:

\begin{align}
    I_- = \mathcal{F}\left(MTF(x)\right) = \mathcal{F}\left(R(-x) \ast E(x)\right) = \mathcal{F}(R(-x)) \cdot \mathcal{F}(E(x))
\end{align}

In the optics context, this is the product of the PSFs from both beams, and therefore if the reference beam is well known, it can be divided out to obtain the desired complex electric field $\mathcal{F}(E(x))$.

We note that if there were a common-path aberration, i.e. a phase shift $\phi_a$ on both $\mathcal{F}(E)$ and $\mathcal{F}(R)$ such that we instead had $\mathcal{F}(E) e^{i\phi_a}$ and $\mathcal{F}(R) e^{i\phi_a}$, this would cancel out due to the complex conjugation and yield the same retrieved mode. However, aberrations in only one arm or the other would not be distinguishable from a component of the unknown beam phase. 

\bibliography{report} 

@ARTICLE{Norris20,
       author = {{Norris}, Barnaby R.~M. and {Wei}, Jin and {Betters}, Christopher H. and {Wong}, Alison and {Leon-Saval}, Sergio G.},
        title = "{An all-photonic focal-plane wavefront sensor}",
      journal = {Nature Communications},
         year = 2020,
        month = oct,
       volume = {11},
          eid = {5335},
        pages = {5335},
          doi = {10.1038/s41467-020-19117-w},
archivePrefix = {arXiv},
       eprint = {2003.05158},
 primaryClass = {astro-ph.IM},
       adsurl = {https://ui.adsabs.harvard.edu/abs/2020NatCo..11.5335N}
}

@ARTICLE{Lin22,
       author = {{Lin}, Jonathan and {Fitzgerald}, Michael P. and {Xin}, Yinzi and {Guyon}, Olivier and {Leon-Saval}, Sergio and {Norris}, Barnaby and {Jovanovic}, Nemanja},
        title = "{Focal-plane wavefront sensing with photonic lanterns: theoretical framework}",
      journal = {Journal of the Optical Society of America B Optical Physics},
         year = 2022,
        month = oct,
       volume = {39},
       number = {10},
        pages = {2643},
          doi = {10.1364/JOSAB.466227},
archivePrefix = {arXiv},
       eprint = {2208.10563},
 primaryClass = {astro-ph.IM},
       adsurl = {https://ui.adsabs.harvard.edu/abs/2022JOSAB..39.2643L}
}

@misc{lightbeam,
       author = {{Lin}, J.},
        title = "{Lightbeam: Simulate light through weakly-guiding waveguides}",
 howpublished = {Astrophysics Source Code Library, record ascl:2102.006},
         year = 2021,
        month = feb,
          eid = {ascl:2102.006},
       adsurl = {https://ui.adsabs.harvard.edu/abs/2021ascl.soft02006L}
}

@inproceedings{Romer2025, title={Broadband photonic lantern transfer matrix characterization for wavefront sensing}, volume={13373}, url={https://www.spiedigitallibrary.org/conference-proceedings-of-spie/13373/133730H/Broadband-photonic-lantern-transfer-matrix-characterization-for-wavefront-sensing/10.1117/12.3043722.full}, DOI={10.1117/12.3043722}, abstractNote={Overcoming the diffraction limit is crucial for advancing imaging capabilities in contemporary science. Conventional imaging methods, reliant on intensity measurements of the optical field across a transverse image plane, are inherently constrained by this limit. However, by measuring additional information about the field, such as its phase and spectral decomposition, it is possible to push this limit. We have developed a Photonic Lantern (PL) imaging device that maximizes the amount of recoverable information of a broadband field. Our PL decomposes the field into a modal basis, converting it into multiple single-mode output channels and dispersing it to obtain a spectrally resolved reading. The lantern’s multimode core collects the light at the focal plane and distributes it among several single-mode fiber ports, acting as a spatial demultiplexer. Obtaining the Transfer Matrix (TM) that describes this conversion of a multimode input beam to multiple single-mode outputs at each wavelength is essential to capitalize on the device’s potential. We demonstrate a procedure to obtain the complex-valued TM of a photonic lantern, allowing the full characterization of the device. Our method uses spatial light modulation to probe the input space with a diverse set of multimode fields and examine the produced spectral response of each output fiber. We then computationally retrieve the matrix that relates these input-output pairs for a given wavelength. The PL characterization is done across a broad spectrum allowing the TM to be obtained simultaneously for a large set of desired wavelengths. This method reliably determines the full-field transfer matrix for a given photonic lantern, enabling numerous applications in imaging and beyond.}, booktitle={Photonic Instrumentation Engineering XII}, publisher={SPIE}, author={Romer, Miguel A. and Batarseh, Ameer B. and Crowe, Tara and Conwell, Robert and Dobias, Caleb and Cruz-Delgado, Daniel and Bandres, Miguel A. and Amezcua-Correa, Rodrigo and Eikenberry, Stephen S.}, year={2025}, month=mar, pages={135–140} }

@ARTICLE{Lin25,
       author = {{Lin}, Jonathan and {Fitzgerald}, Michael P. and {Xin}, Yinzi and {Jung Kim}, Yoo and {Guyon}, Olivier and {Norris}, Barnaby and {Betters}, Christopher and {Leon-Saval}, Sergio and {Ahn}, Kyohoon and {Deo}, Vincent and {Lozi}, Julien and {Vievard}, S{\'e}bastien and {Levinstein}, Daniel and {Sallum}, Steph and {Jovanovic}, Nemanja},
        title = "{Experimental and on-sky demonstration of spectrally dispersed wavefront sensing using a photonic lantern}",
      journal = {Optics Letters},
         year = 2025,
        month = apr,
       volume = {50},
       number = {8},
        pages = {2780},
          doi = {10.1364/OL.551624},
archivePrefix = {arXiv},
       eprint = {2505.00765},
 primaryClass = {astro-ph.IM},
       adsurl = {https://ui.adsabs.harvard.edu/abs/2025OptL...50.2780L}
}

@ARTICLE{XinJATIS,
       author = {{Xin}, Yinzi and {Echeverri}, Daniel and {Jovanovic}, Nemanja and {Mawet}, Dimitri and {Leon-Saval}, Sergio and {Amezcua-Correa}, Rodrigo and {Yerolatsitis}, Stephanos and {Fitzgerald}, Michael P. and {Gatkine}, Pradip and {Kim}, Yoo Jung and {Lin}, Jonathan and {Norris}, Barnaby and {Ruane}, Garreth and {Sallum}, Steph},
        title = "{Laboratory demonstration of a Photonic Lantern Nuller in monochromatic and broadband light}",
      journal = {Journal of Astronomical Telescopes, Instruments, and Systems},
         year = 2024,
        month = apr,
       volume = {10},
          eid = {025001},
        pages = {025001},
          doi = {10.1117/1.JATIS.10.2.025001},
archivePrefix = {arXiv},
       eprint = {2404.01426},
 primaryClass = {astro-ph.IM},
       adsurl = {https://ui.adsabs.harvard.edu/abs/2024JATIS..10b5001X}
}

@ARTICLE{GerardWaveDriver,
       author = {{Gerard}, Benjamin L. and {Geringer-Sameth}, Alex and {Sengupta}, Aditya R. and {Perloff}, Alexx and {Messerley}, Michael and {Sanchez}, Dominic F. and {Waswa}, P. and {Moore}, William and {Cook}, Matthew and {Strang}, Eric and et al.},
        title = "{WaveDriver: a Laser Guide Star AO System for HWO}",
      journal = {arXiv e-prints},
         year = 2025,
        month = sep,
          eid = {arXiv:2509.18643},
        pages = {arXiv:2509.18643},
          doi = {10.48550/arXiv.2509.18643},
archivePrefix = {arXiv},
       eprint = {2509.18643},
 primaryClass = {astro-ph.IM},
       adsurl = {https://ui.adsabs.harvard.edu/abs/2025arXiv250918643G}
}

@ARTICLE{SenguptaOnSky,
       author = {{Sengupta}, Aditya R. and {Diaz}, Jordan and {DeMartino}, Matthew and {Jensen-Clem}, Rebecca and {Cetre}, Sylvain and {Gates}, Elinor and {Bundy}, Kevin and {Dillon}, Daren and {Hinz}, Philip and {Salama}, Ma{\"\i}ssa and et al.},
        title = "{On-sky Demonstration of Second-stage Wave-front Control with a Photonic Lantern}",
      journal = {\aj},
         year = 2026,
        month = feb,
       volume = {171},
       number = {2},
          eid = {65},
        pages = {65},
          doi = {10.3847/1538-3881/ae2617},
archivePrefix = {arXiv},
       eprint = {2511.20560},
 primaryClass = {astro-ph.IM},
       adsurl = {https://ui.adsabs.harvard.edu/abs/2026AJ....171...65S}
}

@ARTICLE{Vievard24,
       author = {{Vievard}, S. and {Lallement}, M. and {Leon-Saval}, S. and {Guyon}, O. and {Jovanovic}, N. and {Huby}, E. and {Lacour}, S. and {Lozi}, J. and {Deo}, V. and {Ahn}, K. and et al.},
        title = "{Spectroscopy using a visible photonic lantern at the Subaru Telescope: Laboratory characterization and the first on-sky demonstration on Ikiiki ({\ensuremath{\alpha}} Leo) and 'Aua ({\ensuremath{\alpha}} Ori)}",
      journal = {\aap},
         year = 2024,
        month = nov,
       volume = {691},
          eid = {A140},
        pages = {A140},
          doi = {10.1051/0004-6361/202450234},
archivePrefix = {arXiv},
       eprint = {2409.06958},
 primaryClass = {astro-ph.IM},
       adsurl = {https://ui.adsabs.harvard.edu/abs/2024A&A...691A.140V}
}

@ARTICLE{Birks15,
       author = {{Birks}, T.~A. and {Gris-S{\'a}nchez}, I. and {Yerolatsitis}, S. and {Leon-Saval}, S.~G. and {Thomson}, R.~R.},
        title = "{The photonic lantern}",
      journal = {Advances in Optics and Photonics},
         year = 2015,
        month = jun,
       volume = {7},
       number = {2},
        pages = {107},
          doi = {10.1364/AOP.7.000107},
archivePrefix = {arXiv},
       eprint = {1503.02837},
 primaryClass = {physics.optics},
       adsurl = {https://ui.adsabs.harvard.edu/abs/2015AdOP....7..107B}
}

@inproceedings{Sengupta25,
author = {Aditya R. Sengupta and Vincent Chambouleyron and Jordan Diaz and Matthew DeMartino and Rebecca Jensen-Clem and Benjamin L. Gerard and Michael J. Messerly and Paul Pax and Daren Dillon and Kevin Bundy and Maria Cuevas and Sylvain Cetre and Bruce Macintosh and Caleb Dobias and Tara Crowe and Stephen S. Eikenberry and Rodrigo Amezcua-Correa and Stephanos Yerolatsitis},
title = {{Experimental validation of photonic lantern imaging and wavefront sensing performance}},
volume = {13627},
booktitle = {Techniques and Instrumentation for Detection of Exoplanets XII},
editor = {Garreth J. Ruane and Maxwell A. Millar-Blanchaer},
organization = {International Society for Optics and Photonics},
publisher = {SPIE},
pages = {136271X},
year = {2025},
doi = {10.1117/12.3064074},
URL = {https://doi.org/10.1117/12.3064074}
}

@ARTICLE{Taras26,
       author = {{Taras}, Adam K. and {Norris}, Barnaby R.~M. and {Betters}, Christopher and {Ross-Adams}, Andrew and {Tuthill}, Peter G. and {Wei}, Jin and {Leon-Saval}, Sergio},
        title = "{Illuminating the lantern: coherent, spectro-polarimetric characterization of a multimode converter}",
      journal = {Optics Express},
         year = 2026,
        month = jan,
       volume = {34},
       number = {1},
        pages = {1012},
          doi = {10.1364/OE.583186},
archivePrefix = {arXiv},
       eprint = {2510.25330},
 primaryClass = {physics.optics},
       adsurl = {https://ui.adsabs.harvard.edu/abs/2026OExpr..34.1012T}
}

@software{cbeam,
       author = {{Lin}, Jonathan W.},
        title = "{cbeam: Coupled-mode propagator for slowly-varying waveguides}",
 howpublished = {Astrophysics Source Code Library, record ascl:2404.001},
         year = 2024,
        month = apr,
          eid = {ascl:2404.001},
archivePrefix = {ascl},
       eprint = {2404.001},
       adsurl = {https://ui.adsabs.harvard.edu/abs/2024ascl.soft04001L}
}

@ARTICLE{LinCoupledModeTheory,
       author = {{Lin}, Jonathan},
        title = "{Coupled-mode theory for astrophotonics}",
      journal = {arXiv e-prints},
         year = 2024,
        month = nov,
          eid = {arXiv:2411.08118},
        pages = {arXiv:2411.08118},
          doi = {10.48550/arXiv.2411.08118},
archivePrefix = {arXiv},
       eprint = {2411.08118},
 primaryClass = {astro-ph.IM},
       adsurl = {https://ui.adsabs.harvard.edu/abs/2024arXiv241108118L}
}

@INPROCEEDINGS{Rypalla24,
       author = {{Rypalla}, Julian and {Vje{\v{s}}nica}, Stella and {Madhav}, Kalaga and {Lorenz}, Adrian and {Eschrich}, Tina and {Schm{\"a}lzlin}, Elmar and {Dinkelaker}, Aline N. and {Roth}, Martin M.},
        title = "{On the large quantity fabrication and reproducibility of all-fiber photonic lanterns}",
    booktitle = {Advances in Optical and Mechanical Technologies for Telescopes and Instrumentation VI},
         year = 2024,
       editor = {{Navarro}, Ram{\'o}n and {Jedamzik}, Ralf},
       series = {Society of Photo-Optical Instrumentation Engineers (SPIE) Conference Series},
       volume = {13100},
        month = aug,
          eid = {131006P},
        pages = {131006P},
          doi = {10.1117/12.3018846},
       adsurl = {https://ui.adsabs.harvard.edu/abs/2024SPIE13100E..6PR}
}

@article{Cuche00,
author = {Etienne Cuche and Pierre Marquet and Christian Depeursinge},
journal = {Appl. Opt.},
number = {23},
pages = {4070--4075},
publisher = {Optica Publishing Group},
title = {Spatial filtering for zero-order and twin-image elimination in digital off-axis holography},
volume = {39},
month = {Aug},
year = {2000},
url = {https://opg.optica.org/ao/abstract.cfm?URI=ao-39-23-4070},
doi = {10.1364/AO.39.004070},
}

@INPROCEEDINGS{Kim24spectralcharacterization,
       author = {{Kim}, Yoo Jung and {Fitzgerald}, Michael P. and {Lin}, Jonathan and {Lozi}, Julien and {Vievard}, Sebastien and {Jovanovic}, Nemanja and {Leon-Saval}, Sergio and {Ahn}, Kyohoon and {Betters}, Christopher and {Deo}, Vincent and et al.},
        title = "{Spectral characterization of 3-port photonic lantern for spectroastrometry}",
    booktitle = {Optical and Infrared Interferometry and Imaging IX},
         year = 2024,
       editor = {{Kammerer}, Jens and {Sallum}, Stephanie and {Sanchez-Bermudez}, Joel},
       series = {Society of Photo-Optical Instrumentation Engineers (SPIE) Conference Series},
       volume = {13095},
        month = aug,
          eid = {130950T},
        pages = {130950T},
          doi = {10.1117/12.3019017},
archivePrefix = {arXiv},
       eprint = {2411.02501},
 primaryClass = {astro-ph.IM},
       adsurl = {https://ui.adsabs.harvard.edu/abs/2024SPIE13095E..0TK}
}

@INPROCEEDINGS{Sengupta24,
       author = {{Sengupta}, Aditya R. and {Diaz}, Jordan and {Gerard}, Benjamin L. and {Jensen-Clem}, Rebecca and {Dillon}, Daren and {DeMartino}, Matthew and {Bundy}, Kevin and {Cetre}, Sylvain and {Chambouleyron}, Vincent},
        title = "{Photonic lantern wavefront reconstruction in a multi-wavefront sensor single-conjugate adaptive optics system}",
    booktitle = {Adaptive Optics Systems IX},
         year = 2024,
       editor = {{Jackson}, Kathryn J. and {Schmidt}, Dirk and {Vernet}, Elise},
       series = {Society of Photo-Optical Instrumentation Engineers (SPIE) Conference Series},
       volume = {13097},
        month = aug,
          eid = {130971J},
        pages = {130971J},
          doi = {10.1117/12.3017873},
       adsurl = {https://ui.adsabs.harvard.edu/abs/2024SPIE13097E..1JS}
}

@ARTICLE{Dobias26,
       author = {{Dobias}, Caleb and {R{\"o}mer}, Miguel A. and {Bhargava}, Swati and {Crowe}, Tara and {Quinn Reyes}, Liza F. and {Smith}, David and {Barzallo}, Matias and {Cruz-Delgado}, Daniel and {Leon-Saval}, Sergio and {Yerolatsitis}, Stephanos and et al.},
        title = "{Wavelength-dependent evolution of full-field transfer matrices in photonic lanterns}",
      journal = {Optics Express},
         year = 2026,
        month = may,
       volume = {34},
       number = {9},
        pages = {17217},
          doi = {10.1364/OE.595908},
archivePrefix = {arXiv},
       eprint = {2604.22091},
 primaryClass = {physics.optics},
       adsurl = {https://ui.adsabs.harvard.edu/abs/2026OExpr..3417217D}
}

@ARTICLE{Galicher10,
       author = {{Galicher}, R. and {Baudoz}, P. and {Rousset}, G. and {Totems}, J. and {Mas}, M.},
        title = "{Self-coherent camera as a focal plane wavefront sensor: simulations}",
      journal = {\aap},
         year = 2010,
        month = jan,
       volume = {509},
          eid = {A31},
        pages = {A31},
          doi = {10.1051/0004-6361/200912902},
archivePrefix = {arXiv},
       eprint = {0911.2465},
 primaryClass = {astro-ph.IM},
       adsurl = {https://ui.adsabs.harvard.edu/abs/2010A&A...509A..31G}
}

@ARTICLE{Gerard18,
       author = {{Gerard}, Benjamin L. and {Marois}, Christian and {Galicher}, Rapha{\"e}l},
        title = "{Fast Coherent Differential Imaging on Ground-based Telescopes Using the Self-coherent Camera}",
      journal = {\aj},
         year = 2018,
        month = sep,
       volume = {156},
       number = {3},
          eid = {106},
        pages = {106},
          doi = {10.3847/1538-3881/aad23e},
archivePrefix = {arXiv},
       eprint = {1806.02881},
 primaryClass = {astro-ph.IM},
       adsurl = {https://ui.adsabs.harvard.edu/abs/2018AJ....156..106G}
}

@ARTICLE{Chambouleyron23,
       author = {{Chambouleyron}, V. and {Fauvarque}, O. and {Plantet}, C. and {Sauvage}, J.-F. and {Levraud}, N. and {Ciss{\'e}}, M. and {Neichel}, B. and {Fusco}, T.},
        title = "{Modeling noise propagation in Fourier-filtering wavefront sensing, fundamental limits, and quantitative comparison}",
      journal = {\aap},
         year = 2023,
        month = feb,
       volume = {670},
          eid = {A153},
        pages = {A153},
          doi = {10.1051/0004-6361/202245351},
archivePrefix = {arXiv},
       eprint = {2212.13577},
 primaryClass = {astro-ph.IM},
       adsurl = {https://ui.adsabs.harvard.edu/abs/2023A&A...670A.153C}
}

@INPROCEEDINGS{Sengupta26,
	author = {{Sengupta}, Aditya R. and {Jensen-Clem}, Rebecca and {Por}, Emiel and {Gerard}, Benjamin L. and {Chambouleyron}, Vincent and {Diaz}, Jordan and {Weber-Porter}, Zoe and {Kim}, Yoo Jung and {Sallum}, Steph and {Bundy}, Kevin and {Gagnebin}, Anna K. and {DeMartino}, Matthew  and {Dillon}, Daren and {Cetre}, Sylvain and {Macintosh}, Bruce and {Dobias}, Caleb and {Crowe}, Tara and {Eikenberry}, Stephen S. and {Amezcua-Correa}, Rodrigo and {Yerolatsitis}, Stephanos},
	title = "{Experimentally-determined performance limits for joint imaging and wavefront sensing with a photonic lantern}",
	booktitle = {These proceedings},
	year = 2026,
	series = {{{SPIE}}},
	volume = {14150-167}
}

@article{Leon-Saval05,
author = {S. G. Leon-Saval and T. A. Birks and N. Y. Joly and A. K. George and W. J. Wadsworth and G. Kakarantzas and P. St.J. Russell},
journal = {Opt. Lett.},
number = {13},
pages = {1629--1631},
publisher = {Optica Publishing Group},
title = {Splice-free interfacing of photonic crystal fibers},
volume = {30},
month = {Jul},
year = {2005},
url = {https://opg.optica.org/ol/abstract.cfm?URI=ol-30-13-1629},
doi = {10.1364/OL.30.001629},
}
\bibliographystyle{spiebib} 

\end{document}